\documentclass[twocolumn,prl,superscriptaddress,showpacs,preprintnumbers,amsmath,amssymb]{revtex4}

\usepackage{graphicx}
\usepackage{dcolumn}
\usepackage{bm}

\begin{document}

\title{Intrinsic vs. extrinsic anomalous Hall effect in ferromagnets}

\author{Shigeki Onoda}
\email{sonoda@appi.t.u-tokyo.ac.jp}
\affiliation{
  Spin Superstructure Project, ERATO, Japan Science and Technology Agency, \\
  c/o Department of Applied Physics, University of Tokyo, Tokyo 113-8656, Japan
}
\author{Naoyuki Sugimoto}%
\affiliation{
  CREST, Department of Applied Physics, University of Tokyo, Tokyo 113-8656, Japan
}%
\author{Naoto Nagaosa}
\affiliation{
  CREST, Department of Applied Physics, University of Tokyo, Tokyo 113-8656, Japan
}%
\affiliation{
  Correlated Electron Research Center, National Institute of Advanced Industrial Science and Technology, Tsukuba, Ibaraki 305-8562, Japan
}

\date{\today}

\begin{abstract}
  A unified theory of the anomalous Hall effect (AHE) is presented 
for multi-band ferromagnetic metallic systems with dilute impurities. 
In the clean limit, the AHE is mostly due to the extrinsic skew-scattering. 
When the Fermi level is located around anti-crossing of band 
dispersions split by spin-orbit interaction, the intrinsic AHE to be 
calculated {\it ab initio} is resonantly enhanced by its non-perturbative 
nature, revealing the extrinsic-to-intrinsic crossover which occurs
when the relaxation rate is comparable to the spin-orbit 
interaction energy.
\end{abstract}

\maketitle

Early experimental works on the Hall effect in ferromagnetic metals 
led a semi-empirical relation of 
the Hall resistivity $\rho_{xy}$ to a {\it weak} applied magnetic field $H$ 
and the spontaneous magnetization $M$ both along the $z$ direction; 
$\rho_{xy} = R_H H + 4 \pi R_s M$ with the normal and the anomalous Hall 
coefficients $R_H$ and $R_s$, respectively~\cite{Hurd}. 
This anomalous Hall effect (AHE)~\cite{Hurd} has been one of the most 
fundamental and intriguing but controversial issues in condensed-matter 
physics~\cite{KarplusLuttinger54,Smit55,Luttinger58,Kondo62,Berger70,Nozieres73}. It has not been clarified yet if the AHE is originated purely from extrinsic scattering or has an intrinsic 
contribution from the electronic band structure, which penetrates even 
recent debates on the interpretation of the 
experiments~\cite{Ohno,Lee_science04,Manyala_nma04}. Theoretically, 
a unified description of both intrinsic and extrinsic contributions is 
called for but has not been considered seriously. 
It even reveals their nontrivial interplay and crossover 
and explains the AHE in a whole region from the clean limit 
to the hopping-conduction region (see Fig.~\ref{fig:scaling}), 
which are main goals of the present study. 

Dissipationless and topological nature of the Hall effect has been highlighted 
by the discovery of the quantum Hall effect~\cite{Prange} in two-dimensional 
electron systems under a {\it strong} magnetic field. In the St\u{r}eda 
formula~\cite{Streda82} of the electric conductivity tensor 
$\sigma_{ij}^{\rm tot}=\sigma_{ij}^I+\sigma_{ij}^{II}$, in ideal cases where the Fermi level is located within the energy gap, 
the Fermi-surface contribution $\sigma_{ij}^I$ vanishes and 
the quantum contribution $\sigma_{ij}^{II}$ yields the TKNN 
formula~\cite{TKNN}
\begin{equation}
  \sigma_{ij}^{TKNN} = -\epsilon_{ij\ell}e^2\hbar
  \sum_n\int\!\frac{d^d\bm{p}}{(2\pi\hbar)^2}b_n^\ell(\bm{p}) f(\varepsilon_n(\bm{p}))
  \label{eq:TKNN}
\end{equation}
with the electronic charge $e$, the Planck constant $h=2\pi\hbar$, the Fermi 
distribution function $f(\varepsilon)$, and the anti-symmetric tensor 
$\epsilon_{ij\ell}$. We have introduced the eigenenergy 
$\varepsilon_n(\bm{p})$, the Berry curvature 
$\bm{b}_n(\bm{p}) = \bm{\nabla}_p\times \bm{a}_n(\bm{p})$ 
and the Berry connection 
$\bm{a}_n(\bm{p}) = i \langle n,{\bm{p}}| \bm{\nabla}_p | n,{\bm{p}}\rangle$ 
of the generalized Bloch wave function $|n,\bm{p}\rangle$ with the band index 
$n$ and the Bloch momentum $\bm{p}$.
Each band has a topological integer called the Chern number
$C_n\equiv -\int\frac{d^2\bm{p}}{(2\pi\hbar)^2}b_n^z(\bm{p})$.
Their summation over the occupied bands determines the integer 
$\nu$ (Chern number) for the quantization 
$\sigma_{xy}^\text{tot}=\nu e^2/h$ in insulators.
Then, adiabatic semi-classical wave-packet equations have been devised to 
incorporate this Berry-curvature effect into the equations of 
motion~\cite{SundaramNiu99}.

Historically, Karplus-Luttinger~\cite{KarplusLuttinger54} initiated an 
intrinsic mechanism of the AHE in a band model for ferromagnetic metals 
with the spin-orbit interaction, which coincides with the TKNN formula 
$\sigma_{xy}^{TKNN}$~\cite{MOnodaNagaosa02,JungwirthNiuMacDonald02}. 
This reflects that the spin-orbit interaction bears 
a nontrivial topological structure in the Bloch wave functions of 
ferromagnets by splitting band dispersions, which originally cross at 
a certain momentum $\bm{p}_0$, with a transfer of Chern numbers among
the bands. This phenomenon called ``parity anomaly'' has a non-perturbative 
nature~\cite{parity}, and points to an importance of the anti-crossing points 
with a small gap $2\Delta_0$, which is identified with the spin-orbit 
interaction energy $\varepsilon_{SO}$.
When the Fermi level is located around such anti-crossing of dispersions, 
as found in recent {\it ab inito} calculations for SrRuO$_3$~\cite{Fang03}
and the bcc Fe~\cite{Yao04}, the magnitude of $\sigma_{xy}^{TKNN}$ is 
resonantly enhanced and approaches $e^2/h=3.87\times10^{-5}\ \Omega^{-1}$ 
in two dimensions and $e^2/ha\sim 10^3 \ \Omega^{-1}\ {\rm cm}^{-1}$ in three 
dimensions with the lattice constant $a\approx4$~\AA~\cite{Fang03,Yao04}.

\begin{figure}
  \begin{center}
    \includegraphics[width=6.0cm]{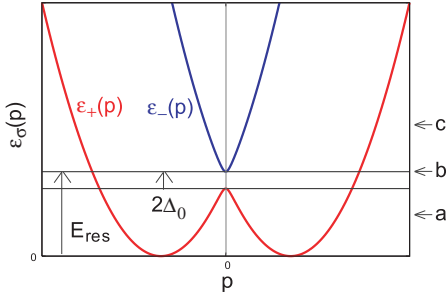}
    \end{center}
  \caption{\label{fig:dispersion}(Color online) Energy dispersions for the Hamiltonian $\hat{H}_0$ in Eq.~(\ref{eq:H}). }
\end{figure}

On the other hand, adiabatic semi-classical Boltzmann transport analyses 
suggest that impurity scattering produces the AHE through the ``skewness'' 
\cite{Smit55,Luttinger58,Nozieres73} or the side 
jump~\cite{Berger70,Nozieres73}. The skew-scattering contribution diverges 
in the clean limit as 
$\sigma_{xy}^{\rm skew}=\sigma_{xx}S=\frac{2e^2}{ha}\frac{E_F\tau}{\hbar}S$ 
with the life time $\tau$ and the Fermi energy $E_F$. Here, 
$S\sim \varepsilon_{SO}v_{\rm imp}/W^2 (\ll1)$ is the skewness factor 
with $W$ being the bandwidth or the inverse of the density of states and 
$v_{\rm imp}$ the impurity potential strength.

Generic model that fully takes into account both the ``parity anomaly'' associated with the anti-crossing of band dispersions and the impurity scattering can be obtained by expanding the Hamiltonian at a fixed $p_z$ with respect to the momentum $\bm{p}$ measured from the originally crossing point $\bm{p}_0$ of two dispersions;
\begin{equation}
\hat{H}_0+\hat{H}_\text{imp}=
-\Delta_0\hat{\sigma}^z+v\bm{p}\cdot\hat{\bm{\sigma}}\times\bm{e}^z+
\frac{\bm{p}^2}{2m}+v_{\rm imp}\sum_{\bm{r}_{\rm imp}}\delta(\bm{r}-\bm{r}_{\rm imp})
\label{eq:H}
\end{equation}
with the position $\bm{r}$ of electron, the Pauli matrices 
$\hat{\bm{\sigma}}=(\hat{\sigma}_x,\hat{\sigma}_y,\hat{\sigma}_z)$, 
the unit vector $\bm{e}^z$ in the $z$ direction, and an impurity at a 
position $\bm{r}_{\rm imp}$.
The first term corresponds to the level splitting $2\Delta_0=\varepsilon_{SO}$ 
of two bands at the anti-crossing momentum. 
The second term gives the linear dispersion with the velocity $v$. 
The third term represents the quadratic dispersion with an effective mass $m$, 
whose anisotropy has been neglected since it is unimportant. 
This model has two band dispersions $\varepsilon_{\pm,\bm{p}}$ 
as shown in Fig.~\ref{fig:dispersion}. 
Henceforth, the bottom of the lower band is chosen 
as the origin of the energy and the bottom of the upper 
band denoted as $E_{\rm res}=\varepsilon_-(0)$ is taken as an energy unit. 
The model possesses the gauge flux 
$b^z_{\pm,\bm{p}} =\mp\lambda^2\Delta_0/2\Delta_{\bm{p}}^3$ with 
$\Delta_{\bm{p}}=\sqrt{\lambda^2 \bm{p}^2
+\Delta_0^2}$~\cite{Dugaev05,Sinitsyn05}. In the case of resonance, 
$E_F \in [E_{\rm res}-2\Delta_0,E_{\rm res}]$, $\sigma_{xy}^{TKNN}$ 
approaches the maximum value $e^2/2h$. Away from this resonance, dominant 
contributions from the momentum region around $\bm{p}=0$ cancel out each 
other or do not appear, leading to a suppression of 
$\sigma_{xy}^{TKNN}(\approx(e^2/h)(\varepsilon_{SO}/E_F))$, 
where the perturbation expansion in $\varepsilon_{SO}$ is justified. Therefore, the present model, Eq.~(\ref{eq:H}), can be regarded as a generic continuum model for a momentum region that gives a major contribution to the AHE.
By definition of the 
anti-crossing, $\Delta_0$ does not change its sign as a function of $p_z$. 
This removes a concern that the integration over $p_z$ might lead to a 
cancellation. 

We employ the Keldysh formalism for non-equilibrium Green's functions, which 
has recently been reformulated for generic multi-component 
systems~\cite{Onoda06}. We consider Green's functions and 
self-energies under an applied constant electric field 
$\bm{E}=(0,E_y)$; $\hat{G}^\alpha(\varepsilon,\bm{p})$ and 
$\hat{\Sigma}^\alpha(\varepsilon)$ with $\alpha=R,A,<$ for 
the retarded, advanced and lesser components, respectively. 
$\varepsilon$ and $\bm{p}$ represent the covariant energy and momentum 
in the Wigner representation~\cite{Onoda06}. 
$\hat{G}^\alpha(\varepsilon,\bm{p})$ and 
$\hat{\Sigma}^\alpha(\varepsilon)$ can be expanded in $E_y$ as
\begin{eqnarray}
  \hat{G}^\alpha(\varepsilon,\bm{p}) &=& \hat{G}^\alpha_0(\varepsilon,\bm{p}) 
+ e\hbar E_y \hat{G}^\alpha_{E_y}(\varepsilon,\bm{p})+O(E_y^2),
  \label{eq:g}\\
  \hat{\Sigma}^\alpha(\varepsilon) &=& \hat{\Sigma}^\alpha_0(\varepsilon) 
+ e\hbar E_y \hat{\Sigma}^\alpha_{E_y}(\varepsilon)+O(E_y^2).
  \label{eq:Sigma}
\end{eqnarray}
Henceforth, functionals with the subscripts 0 and $E_y$ denote those in the 
absence of and the gauge-invariant linear response to $E_y$. 
Due to the $\delta$-functional form of the impurity potential, the 
self-energies are local. $\hat{G}^{R,A}_0$ satisfies the familiar 
Dyson equation,
\begin{equation}
  \hat{G}^{R,A}_0(\varepsilon,\bm{p})=[\varepsilon-\hat{H}_0(\bm{p})
-\hat{\Sigma}^{R,A}_0(\varepsilon)]^{-1}.
  \label{eq:G^R,A:0}
\end{equation}
The self-consistent equations for $\hat{G}^{R,A,<}_{E_y}$ are obtained by 
expanding the Dyson equation in the electric field~\cite{Onoda06}. 
It is convenient to decompose $\hat{G}^<_{E_y}$ and $\hat{\Sigma}^<_{E_y}$ 
into two;
\begin{eqnarray}
  \hat{G}^<_{E_y}(\varepsilon,\bm{p})&=&
\hat{G}^<_{E_y,I}(\varepsilon,\bm{p})\partial_\varepsilon 
f(\varepsilon)+\hat{G}^<_{E_y,II}(\varepsilon,\bm{p})f(\varepsilon)
  \label{eq:G^<:E}\\
  \hat{\Sigma}^<_{E_y}(\varepsilon)&=&\hat{\Sigma}^<_{E_y,I}(\varepsilon)
\partial_\varepsilon f(\varepsilon)+\hat{\Sigma}^<_{E_y,II}(\varepsilon)
f(\varepsilon)
  \label{eq:Sigma^<:E}\\
  \hat{G}^<_{E_y,II}(\varepsilon,\bm{p})&=&\hat{G}^A_{E_y}(\varepsilon,\bm{p})
-\hat{G}^R_{E_y}(\varepsilon,\bm{p})
  \label{eq:G^<:E,II}\\
  \hat{\Sigma}^<_{E_y,II}(\varepsilon)&=&\hat{\Sigma}^A_{E_y}(\varepsilon)
-\hat{\Sigma}^R_{E_y}(\varepsilon).
  \label{eq:Sigma^<:E,II}
\end{eqnarray}
$\hat{G}^<_{E_y,I}$ and $\hat{\Sigma}^<_{E_y,I}$ can be self-consistently 
determined from the quantum Boltzmann equation in the first order in $E_y$,
\begin{eqnarray}
  \lefteqn{\left[\hat{G}^<_{E_y,I},\hat{H}_0\right]+\hat{G}^<_{E_y,I}
\hat{\Sigma}^A_0-\hat{\Sigma}^R_0\hat{G}^<_{E_y,I}}
  \nonumber\\
  &&=\hat{\Sigma}^<_{E_y,I}\hat{G}^A_0-\hat{G}^R_0\hat{\Sigma}^<_{E_y,I}
  -\frac{i}{2}\left[\hat{v}_y,\hat{G}^A_0-\hat{G}^R_0\right]_+
  \nonumber\\
  &&+\frac{i}{2}\left(
  (\hat{\Sigma}^A_0-\hat{\Sigma}^R_0)(\partial_{p_y}\hat{G}^A_0)
  +(\partial_{p_y}\hat{G}^R_0)(\hat{\Sigma}^A_0-\hat{\Sigma}^R_0)\right)
  \label{eq:G^<:E,I}
\end{eqnarray}
with the velocity $\hat{v}_i(\bm{p})=\partial_{p_i}\hat{H}_0(\bm{p})$, while 
$\hat{G}^{R,A}_{E_y}$ and $\hat{\Sigma}^{R,A}_{E_y}$ are determined from 
the other self-consistent equation,
\begin{eqnarray}
  \lefteqn{\hat{G}^{R,A}_{E_y}
  =\hat{G}^{R,A}_0\hat{\Sigma}^{R,A}_{E_y}\hat{G}^{R,A}_0}
  \nonumber\\
  &&{}-\frac{i}{2}
  \left(\hat{G}^{R,A}_0\hat{v}_y(\partial_\varepsilon\hat{G}^{R,A}_0)
-(\partial_\varepsilon\hat{G}^{R,A}_0)\hat{v}_y\hat{G}^{R,A}_0\right).
  \label{eq:G^R,A:E}
\end{eqnarray}

We can exactly calculate the self-energies $\hat{\Sigma}^{R,A}_0$ and 
$\hat{\Sigma}^{R,A,<}_{E_y}$ up to the $n_{\rm imp}$-linear terms by means 
of the $T$-matrix approximation;
\begin{eqnarray}
  \hat{\Sigma}^{R,A}_0(\varepsilon)&=&n_{\rm{imp}}\hat{T}^{R,A}_0(\varepsilon)
  \label{eq:Sigma^R,A:0}\\
  \hat{T}^{R,A}_0(\varepsilon)&=&v_{\rm{imp}}\left(1-v_{\rm imp}\int
\frac{d^2\bm{p}}{(2\pi\hbar)^2}\hat{G}^{R,A}_0(\varepsilon,\bm{p})\right)^{-1}
  \label{eq:g^R,A:0}
\end{eqnarray}
for the zeroth-order in $E_y$ and
\begin{eqnarray}
  \hat{\Sigma}^<_{E_y,I}(\varepsilon)\!\!&=&\!\!n_{\rm{imp}}
\hat{T}^R_0(\varepsilon)\!\!\int\!\!\frac{d^2\bm{p}}{(2\pi\hbar)^2}
\hat{G}^<_{E_y,I}(\varepsilon,\bm{p})\hat{T}^A_0(\varepsilon)
  \label{eq:Sigma^<:E,I}\\
  \hat{\Sigma}^{R,A}_{E_y}(\varepsilon)\!\!&=&\!\!n_{\rm imp}
\hat{T}^{R,A}_0(\varepsilon)\!\!\int\!\!\frac{d^2\bm{p}}{(2\pi\hbar)^2}
\hat{G}^{R,A}_{E_y}(\varepsilon,\bm{p})\hat{T}^{R,A}_0(\varepsilon)
  \label{eq:Sigma^R,A:E}
\end{eqnarray}
for the first-order in $E_y$.
We solve Eqs.~(\ref{eq:G^R,A:0}), (\ref{eq:Sigma^R,A:0}) and 
(\ref{eq:g^R,A:0}) self-consistently to obtain $\hat{G}^{R,A}_0$ and 
$\hat{\Sigma}^{R,A}_0$. Next, they are plugged into Eqs.~(\ref{eq:G^<:E,I}) 
and (\ref{eq:Sigma^<:E,I}) to solve $\hat{G}^<_{E_y,I}$ and 
$\hat{\Sigma}^<_{E_y,I}$ self-consistently. 
$\hat{G}^{R,A}_{E_y}$ and $\hat{\Sigma}^{R,A}_{E_y}$  are obtained 
from Eqs.~(\ref{eq:G^R,A:E}) and (\ref{eq:Sigma^R,A:E}) , and hence
$\hat{G}^<_{E_y,II}$ through Eq.~(\ref{eq:G^<:E,II}). 
The conductivity tensors are calculated from  
\begin{eqnarray}
  \sigma_{ij}^I\!\!&=&\!\!-\frac{e^2\hbar}{2\pi i}
  \int\!\!\frac{d^2\bm{p}}{(2\pi\hbar)^2}{\rm Tr}\!
  \left[\hat{v}_i(\bm{p}) \hat{G}^<_{E_j,I}(\varepsilon,\bm{p})\right],
  \label{eq:sigma_I}\\
  \sigma_{ij}^{II}\!\!&=&\!\!e^2\hbar\int\!\!\frac{d\varepsilon}{2\pi i}
  \int\!\!\frac{d^2\bm{p}}{(2\pi\hbar)^2}{\rm Tr}\!
  \left[\hat{v}_i(\bm{p}) \hat{G}^<_{E_j,II}(\varepsilon,\bm{p})\right]\!f(\varepsilon),
  \label{eq:sigma_II}
\end{eqnarray}
with $i,j=x,y$. Note that Eqs.~(\ref{eq:sigma_I}) and (\ref{eq:sigma_II}) 
are along the same spirit as the St\u{r}eda formula~\cite{Streda82}: 
this approach provides the diagrammatic treatment for the St\u{r}eda 
formula~\cite{Onoda06}. Effects of the vertex corrections to 
$\hat{G}^<_{E_j,II}$ cancel each other, 
and hence we can regard $\sigma_{xy}^{II}$ as an intrinsic contribution.
By contrast, the Fermi-surface contribution $\sigma_{xy}^I$ suffers 
from a vertex correction. 
While the intra-band matrix elements correspond to the conventional 
description of both $\sigma_{xx}$ and $\sigma_{xy}$ based on the scattering 
events, the inter-band ones contain a finite intrinsic contribution 
to the AHE as a part of the Berry-curvature term~\cite{Sinitsyn06} and is generically expressed as
\begin{eqnarray}
  \lefteqn{\sigma^{I\ {\rm int}}_{ij}=-\epsilon_{ij\ell}\frac{e^2\hbar}{2}\int\!\frac{d\bm{p}}{(2\pi\hbar)^d}\sum_{n,n'}(\varepsilon_n(\bm{p})-\varepsilon_{n'}(\bm{p}))}
  \nonumber\\
  &&\!\!\!\times\partial_\varepsilon f(\varepsilon_n(\bm{p})){\rm Im}\left(\langle n\bm{p}|\bm{\nabla}_p|n'\bm{p}\rangle\times\langle n'\bm{p}|\bm{\nabla}_p|n\bm{p}\rangle\right)_\ell.
  \label{eq:sigma^I:intrinsic}
\end{eqnarray}

\begin{figure}
  \begin{center}
    \includegraphics[width=7.0cm]{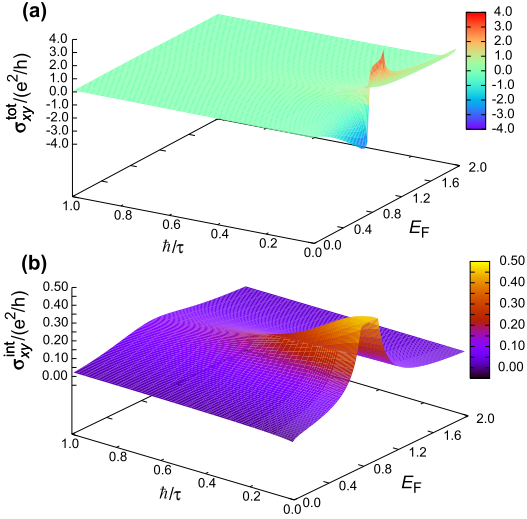}
  \end{center}
  \caption{\label{fig:theory}(Color) (a) The total anomalous Hall conductivity $\sigma_{xy}^{\rm tot}$ against $E_F$ and $\hbar/\tau$ in an energy unit of $E_{\rm res}=1.0$. (b) The intrinsic contribution $\sigma_{xy}^{\rm int}$ for the same parameter values. Note the difference of the scales for $\sigma_{xy}$ in (a) and (b).}
\end{figure}
Figure~\ref{fig:theory}(a) shows the numerical results on 
$\sigma_{xy}^{\rm tot}=\sigma_{xy}^I+\sigma_{xy}^{II}$ against the Fermi 
energy $E_F$ and the Born scattering amplitude 
$\hbar/\tau\equiv n_{\rm imp}v_{\rm imp}^2 m$ 
for a typical set of parameters, $\Delta_0=0.1$, $2mv_{\rm imp}=0.6$, $2m\lambda^2=3.59$, and the energy cutoff is taken as $E_c=3.0$ in an energy unit of $E_{\rm res}=1.0$~\cite{Inoue06}. 
In the clean limit $\hbar/\tau\to0$, $\sigma_{xy}^{\rm tot}$ diverges 
in accordance with the skew-scattering scenario. Strength of the
divergence is proportional to $E_F$ in the low electron-density limit, 
and the sign is inverted around $E_F=\varepsilon_+(0)=E_{\rm res}-2\Delta_0$. 
Sign of the skew-scattering contribution also changes by the sign change of $v_{\rm imp}$.
Figure~\ref{fig:theory}(b) shows the intrinsic contribution 
$\sigma_{xy}^{\rm int}$ calculated by imposing 
$\hat{\Sigma}^{R,A,E}_{E_y}=0$ for the same set of parameters. 
Under the resonant condition for $E_F$, $\sigma_{xy}^{\rm int}$ becomes of the order of $e^2/2h$. With increasing $\hbar/\tau$, it gradually decreases solely due to damping of quasi-particles.

\begin{figure}
  \begin{center}
    \includegraphics[width=7.0cm]{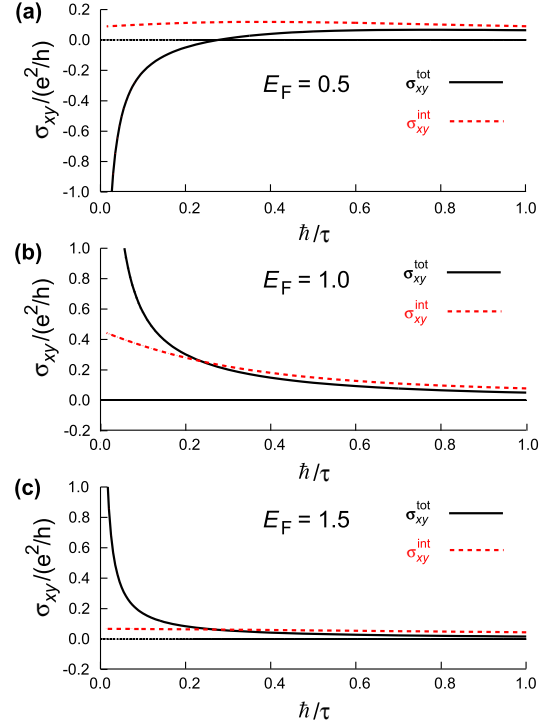}
  \end{center}
  \caption{\label{fig:theory2}(Color online) $\sigma_{xy}^\text{tot}$ and $\sigma_{xy}^\text{int}$ as a function of $\hbar/\tau$ for the same parameter values as Fig.~\ref{fig:theory} with $E_F=0.5$, $1.0$ and $1.5$ for (a), (b), and (c), respectively.}
\end{figure}

By contrast, with increasing $\hbar/\tau$, the extrinsic skew-scattering contribution rapidly decays (Fig.~\ref{fig:theory2}), reflecting that it originates purely from intra-band processes and hence the skewness factor $S$ remains of the order of $\varepsilon_{SO}v_{\rm imp}/W^2$. In moderately dirty cases, the total conductivity nearly merges into the intrinsic value. Namely, there appears a crossover from the extrinsic regime to the intrinsic as $\hbar/\tau$ increases. Especially in the resonant case ($E_F=E_\text{res}$) shown in Fig.~\ref{fig:theory2} (b), the intrinsic contribution is significantly enhanced and the extrinsic-to-intrinsic crossover occurs at $\hbar/\tau\sim\varepsilon_{SO}$. For a small ratio of $\varepsilon_{SO}/E_F\sim 10^{-3}-10^{-2}$~\cite{Fang03,Yao04}, dominance of the intirnsic AHE is realized within the usual clean metal. In reality, the total Hall conductivity is the sum of the contributions from all over the Brillouin zone. Since skew-scattering contributions from other momentum regions are always subject to a similar rapid decay, the above extrinsic-to-intrinsic crossover still occurs unless contributions from all the anti-crossing regions of band dispersions are mutually canceled out by accident.

\begin{figure}
  \begin{center}
    \includegraphics[width=7.0cm]{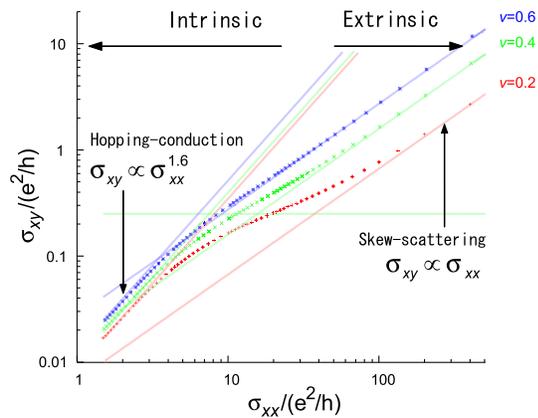}
  \end{center}
  \caption{\label{fig:scaling} (Color online) Scaling plot of $\sigma_{xy}$ versus $\sigma_{xx}$ for the same sets of parameter values as in Fig.~\ref{fig:theory2}(b) except $2mv_{\rm imp}$.
}
\end{figure}

Figure~\ref{fig:scaling} shows the logarithmic plot of $\sigma_{xy}$ against $\sigma_{xx}$ for the same set of parameters as Fig.~\ref{fig:theory2}(b) except for the impurity scattering strength $v_{\rm imp}$, which is varied for different curves. In the clean limit, the curves nicely follow $\sigma_{xy}\propto\sigma_{xx}$ and the ratio $\sigma_{xy}/\sigma_{xx}$ is proportional to $v_{\rm imp}$ for a fixed $\tau$. 
As $\sigma_{xx}=2(e^2/h)E_F\tau/\hbar$ decreases with decrease in $\tau$, the relation exhibits a upward deviation from the linear one, signalling the crossover to the intrinsic regime. A smaller impurity potential strength $v_{\rm imp}$ enlarges the region of the constant behaviour of $\sigma_{xy}$. (Note that we change $n_{\rm imp}$ to control $\hbar/\tau$.) 
Careful experiments are required to test the prediction of the crossover at low temperatures. The magnitude of $\sigma_{xy}$ in the intrinsic regime is consistent with experimentally observed values $\sigma_{xy} \cong 10^2-10^3 \ \Omega^{-1}{\rm cm}^{-1}$ in a $\sigma_{xy}$-constant region of Fe- and Ni-based dilute alloys~\cite{Hurd},
and SrRuO$_3$ and metallic foils~\cite{Asamitsu}. Further decrease of $\tau$ again changes the scaling behaviour to $\sigma_{xy}\propto\sigma_{xx}^{1.6}$, which almost agrees with recent experiments on a disordered pyrochlore ferromagnet Nd$_2$(Mo$_{1-x}$Nb$_x$)$_2$O$_7$~\cite{Iguchi} and on La$_{1-x}$Sr$_x$CoO$_3$~\cite{Asamitsu}. This exponent approximates to the value expected for the normal Hall effect in the hopping conduction regime~\cite{PryadkoAuerbach04}.

Now the source of the confusion over decades is clear. The skew-scattering contribution, though it is rather sensitive to details of the impurity potential and band structure, can be larger than $e^2/h$ in the superclean case $\hbar/\tau\ll\varepsilon_{SO}$, but decays for $\varepsilon_{SO} < \hbar/\tau$. The side-jump contribution is smaller and of the order of $(e^2/h)(\varepsilon_{SO}/E_F)$~\cite{Nozieres73}. Therefore, the intrinsic one, which is of the order of $e^2/h$ under the resonant condition, is dominant over a wide range of the scattering strength $\varepsilon_{SO}<\hbar/\tau< E_F$ (clean case). Although Luttinger reconsidered the Karplus-Luttinger theory~\cite{KarplusLuttinger54} and gave an expansion of $\sigma_{xy}$ in $v_{\rm imp}$, including the skew-scattering contribution as well~\cite{Luttinger58}, it fails to reveal the above crossover in the space of $E_F$, $\varepsilon_{SO}$ and $\hbar/\tau$.

In conclusions, we have shown that the AHE is determined by the intrinsic mechanism when (i) the Fermi level is located around an anti-crossing of band dispersions in the momentum space, (ii) consequently the magnitude of $\sigma_{xy}$ is comparable to $e^2/(ha) \sim 10^3 \ \Omega^{-1}$ cm$^{-1}$, and (iii) the resistivity $\rho_{xx}$ is larger than $ (ha/e^2)(\varepsilon_{SO}/E_F)\sim 1$-$10\ \mu\Omega\ {\rm cm}$. With these conditions, first-principle calculation can give an accurate prediction of $\sigma_{xy}$. The present work resolves a long standing controversy on the mechanism of the AHE in a whole region.

The authors are grateful to A. Asamitsu, T. Miyazato, S. Iguchi, R. Mathieu, and Y. Tokura for fruitful discussion and showing unpublished experimental data, and H. Fukuyama, A. H. MacDonald, J. Sinova, J. Inoue, S. Murakami, K. Nomura, and M. Onoda for stimulating discussion. Numerics was performed using the supercomputer at the Institute of Solid State Physics, University of Tokyo. The work was supported by Grant-in-Aids and NAREGI Nanoscience Project from the Ministry of Education, Culture, Sports, Science, and Technology.

\end{document}